\documentclass[fleqn,usenatbib]{mnras}

\usepackage{newtxtext,newtxmath}

\usepackage[T1]{fontenc}

\DeclareRobustCommand{\VAN}[3]{#2}
\let\VANthebibliography\thebibliography
\def\thebibliography{\DeclareRobustCommand{\VAN}[3]{##3}\VANthebibliography}

\usepackage{graphicx}	
\usepackage{amsmath}	
\usepackage{subcaption}
\usepackage{hyperref}
\usepackage{float}
\usepackage{gensymb}

\title[Broadband atmospheric dispersion corrector]{A new broadband atmospheric dispersion corrector for HROS-TMT}

\author[Bestha et al.]{Manjunath Bestha,$^{1,}$$^{2}$\thanks{E-mail: bestha95@gmail.com} 
Thirupathi Sivarani,$^{1}$
Bachar Wehbe,$^{3,}$$^{4}$
Amirul Hasan,$^{1,}$$^{5}$
Bharat Chandra P,$^{1}$
\newauthor
Devika K Divakar,$^{6}$
Athira Unni,$^{7}$
Parvathy Menon,$^{1,}$$^{2}$
Arun Surya,$^{1}$
Pallavi Saraf $^{1,}$$^{8}$\\
$^{1}$Indian Institute of Astrophysics, Bangalore, India\\
$^{2}$Department of Applied Optics and Photonics, University of
Calcutta, Kolkata, India\\
$^{3}$ Instituto de Astrofísica e Ciências do Espaço, Universidade de Lisboa, Campo Grande, 1749-016 Lisboa, Portugal\\
$^{4}$\ Departamento de Física, Faculdade de Ciências, Universidade de Lisboa, Campo Grande, 1749-016 Lisboa, Portugal\\
$^{5}$ CHRIST (Deemed to be University), Bangalore, India\\
$^{6}$University of Texas, Austin, USA\\
$^{7}$University of California, Santa Cruz, USA\\
$^{8}$\ Physical Research Laboratory, Ahmadabad, India}

\date{Accepted XXX. Received YYY; in original form ZZZ}

\pubyear{\the\year{}}

\begin{document}
\label{firstpage}
\pagerange{\pageref{firstpage}--\pageref{lastpage}}
\maketitle

\begin{abstract}
Atmospheric dispersion causes light from celestial objects with different wavelengths to refract at varying angles as it passes through Earth's atmosphere. This effect results in an elongated image at the focal plane of a telescope and diminishes fiber coupling efficiency into spectrographs.
We propose an optical design that incorporates a Rotational Atmospheric Dispersion Corrector (RADC) to address the broadband dispersion encountered in the multi-object mode of the High-Resolution Optical Spectrograph (HROS) on the Thirty Meter Telescope (TMT).
The RADC corrects the dispersion across the entire wavelength range (0.31-1$~\mu\text{m}$), using Amici prisms optimized for over $90\%$  transmission efficiency and minimal angular deviation of the beam from the optical axis after dispersion correction. For enhanced accuracy, particularly in the blue region, we have, for the first time, implemented the Filippenko (1982) model in Zemax via a custom Dynamic-Link Library (DLL) file.
\end{abstract}



\color{black}
\section{Introduction}

Broadband observations play a key role in astronomical spectroscopic studies; for instance, they help study the detailed chemical
abundances of astronomical objects in a single observation, such as
stellar archaeology and exoplanet atmosphere characterization \citep{Sneden.2000,Hill.2002,sivarani,birkby2018exoplanetatmosphereshighspectral,hd209458b,Pallavi,Avrajit,Rukdee2024,mo_hrts}. However, these observations from the ground are affected by atmospheric dispersion. This phenomenon occurs when light from celestial objects passes through Earth's atmosphere and gets refracted. This leads to elongated images on the focal plane of telescopes \citep{Wynne}. For spectroscopic observations, it causes wavelength-dependent input losses at slit or fiber and reduces the overall spectrograph efficiency.

In slit-based spectrographs, slit losses can be minimized by aligning the slit along the parallactic angle \citep{paralactic}. One effective method to mitigate dispersion in fiber-fed spectrographs is to feed the pupil image into the fibers. However, feeding the pupil image directly into the fiber in crowded fields can allow multiple objects to enter a single fiber. Therefore, dispersion correctors are extensively used in fiber-fed spectrographs to address this.

Upcoming large telescopes, such as the Extremely Large Telescope (ELT)\footnote{\url{https://elt.eso.org/}} \citep{ELT, Padovani_2023} and the Thirty Meter Telescope (TMT)\footnote{\url{https://www.tmt.org/}} \citep{TMT1}, face challenges in correcting atmospheric dispersion across the wide field of view, primarily due to the need for large correction optics. To address this challenge, a segmented Atmospheric Dispersion Corrector (ADC) design concept was proposed for the Large Sky Area Multi-Object Fibre Spectroscopic Telescope (LAMOST), with diameters of $1.75$~m and $1.22$~m for the North and South telescopes, respectively \citep{Lamost_ADC}. However, constructing and maintaining segmented ADCs that align with the curvature of the telescope's focal plane becomes challenging \footnote{Similar to the increase in design complexity of segmented mirror telescopes, as the primary aperture size increases.} as the field of view increases.
 
Therefore, this paper presents an ADC design based on a widely utilized design, tailored for single-object dispersion correction within individual narrow fields. The proposed ADC is intended to provide dispersion-corrected light to the High-Resolution Optical Spectrograph (HROS) \citep{Sivarani_2022}, a second-generation instrument proposed for the TMT. HROS covers a wavelength range from 0.31$~\mu\text{m}$ to 1$~\mu\text{m}$ and supports various operational modes, including Multi-Object Spectroscopy (MOS). In MOS mode, up to six objects can be observed simultaneously, utilizing $1\arcsec{}$ fibers to achieve a spectral resolution of $R = 25000$. By blocking subsequent echelle orders, the slit can accommodate up to forty objects \citep{Sivarani_2022, Manju_ADC}. The ADC is integrated within the fiber positioners responsible for feeding light to the spectrograph \citep{zebri_2014,Manju_Positioner}. The details of TMT and HROS are discussed in Section \ref{overview}.

Atmospheric dispersion correction is typically achieved using one of two common types of correctors in astronomical observations: Linear Atmospheric Dispersion Correctors (LADCs) \citep{Avila_LADC} and Rotational Atmospheric Dispersion Correctors (RADCs) \citep{Wynne_RADC}. An LADC operates by translating one of two identical, oppositely oriented prisms along the optical axis within a converging beam to counteract atmospheric dispersion. However, LADCs introduce lateral shift in the beam that changes with zenith angle, which can lead to throughput losses, especially in systems with stationary fiber inputs \citep{Bahrami_11,app11146261}.
To overcome this limitation, the RADC design, featuring counter-rotating Amici prisms, is incorporated into the fiber positioner, achieving simultaneous correction of both dispersion and beam deviation, thereby enhancing spectrograph efficiency. In this context, the term “beam” refers to the polychromatic light with a reference wavelength of 0.45$~\mu$m.

\label{Introduction}

\section{Overview of the Thirty Meter Telescope and High-Resolution Optical Spectrograph}
\label{overview}
\begin{figure*} 
    \centering
    \includegraphics[width=\textwidth]{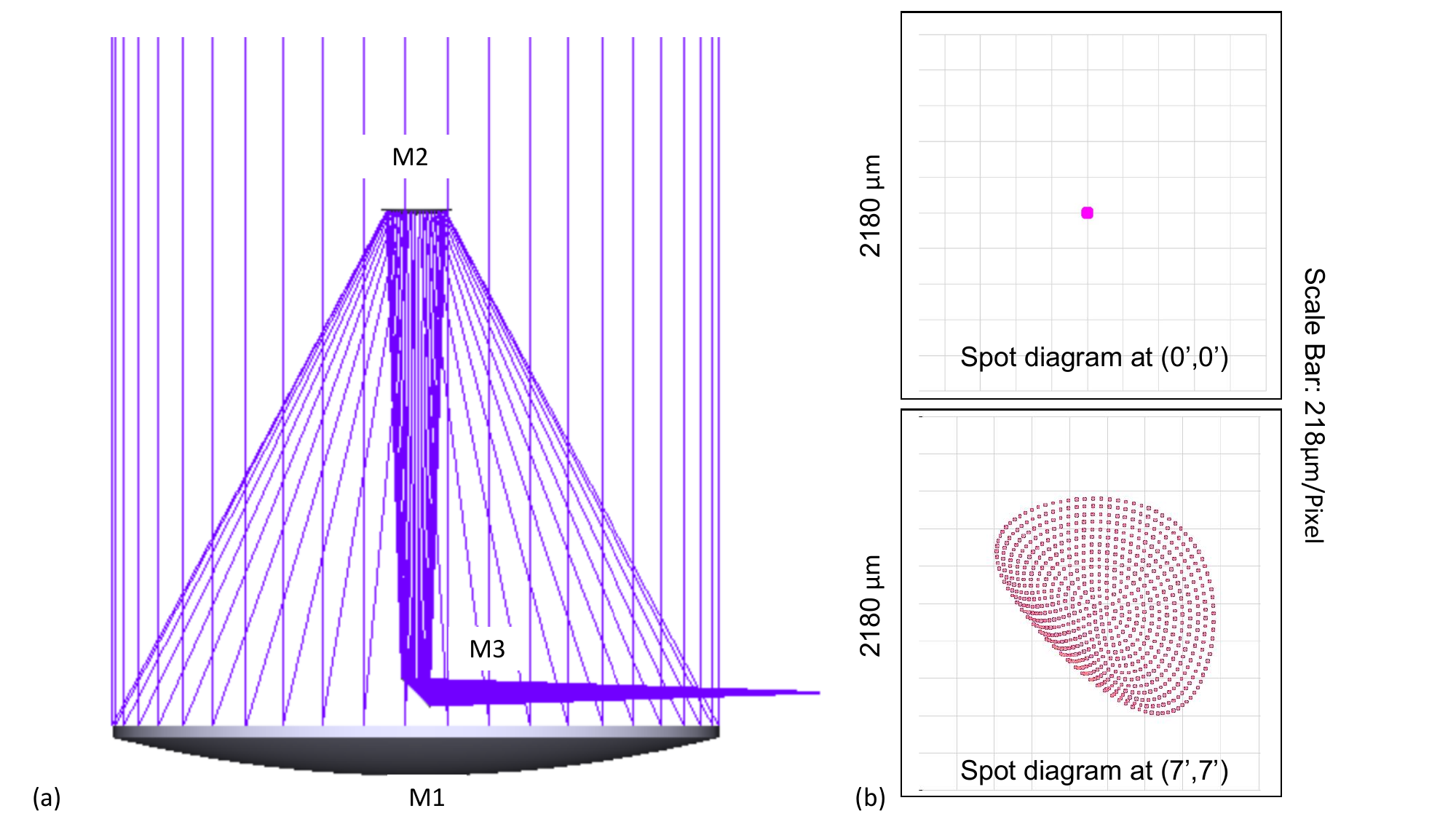}
    \caption{(a) The TMT optical layout, and (b) The spot diagrams for a reference wavelength of 0.45$~\mu\text{m}$. The top panel shows the spot diagram for the on-axis position, while the bottom panel displays the spot diagram for the off-axis position.}

    \label{TMT_Design}
\end{figure*}

The Thirty Meter Telescope (TMT) \citep{TMT1,Sivarani_2022,Manju_ADC} is an alt-azimuth Ritchey–Chrétien (RC) telescope planned for construction on Mauna Kea, Hawaii. It consists of a concave hyperboloid primary mirror (M1), a convex hyperboloid secondary mirror (M2), and a flat tertiary mirror (M3). The primary mirror has a diameter of $30$ meters, which is achieved by assembling $492$ hexagonal segments. The secondary mirror has a diameter of $3.1$ meters (see Table \ref{tmttable}). At the final focus, the telescope delivers a f/15 beam with a plate scale of $2.18\,\text{mm/1$\arcsec{}$}$ and a focal plane diameter of $2.616$ meters. The focal plane has a radius of curvature of $3.01$ meters, which results in a non-telecentric beam. On-axis, TMT achieves an image quality with a geometric radius of $\sim20~\mu$m. However, due to astigmatism, this degrades to $\sim730~\mu$m at an off-axis field point located at $(7\arcmin{}, 7\arcmin{})$, since Ritchey–Chrétien (RC) telescopes are not inherently corrected for astigmatism \citep{RC_Telescope}. The effect is illustrated by the TMT focal plane spot diagram (Figure~\ref{TMT_Design}). The off-axis point corresponds to a radial distance of $\approx$ $9.9\arcmin{}$ from the field center, yielding a total field of view of about $20\arcmin{}$ in diameter.

The TMT is designed to host both adaptive optics-assisted instruments and seeing-limited instruments mounted on either side of the Nasmyth platform. One of the second-generation instruments, the High-Resolution Optical Spectrograph (HROS) \citep{Sivarani_2022}, is seeing-limited and aims to leverage TMT's large aperture even under moderate seeing conditions. It supports operation using direct slit-fed and fiber-fed modes, offering three resolutions: high-resolution ($R = 100{,}000$), standard mode ($R = 50{,}000$), and multi-object spectroscopy (MOS) mode ($R = 25{,}000$), with a wavelength coverage of 0.31--1.0$~\mu$m \citep{Sivarani_2022, Manju_ADC}.

The MOS mode utilizes fiber positioners to collect the light from the focal plane to feed the spectrograph.
The design of the positioner incorporates an ADC along with fore optics, comprising a collimator, a camera, a dichroic, and a pick-off mirror. This integrated assembly is called the Atmospheric Dispersion Correction System (ADCS), and each positioner arm has its own ADCS. 

The pick-off mirror redirects incoming light toward the collimator using small tip and tilt adjustments. For on-axis targets, the mirror is set at a $45^\circ$ angle to the chief ray. For off-axis targets, the angle is adjusted by $\pm \Delta\theta$, where $\Delta\theta$ is the offset between the on-axis and off-axis directions \citep{Manju_Positioner}. The redirected light then passes through the ADC, which corrects for atmospheric dispersion, and is split into two wavelength bands using a dichroic: 0.31--0.45$~\mu$m (blue channel) and 0.44--1.0$~\mu$m (red channel), as listed in Table \ref{tmttable}. A 0.01~$~\mu$m overlap is included in the red channel since the typical dichroic cannot split the light exactly at the desired boundary. The light is then coupled into optical fibres using micro lenses.

\vspace{20pt} 

\begin{figure*}
    \centering
    \includegraphics[width=1\textwidth,trim=0pt 70pt 0pt 70pt,clip]{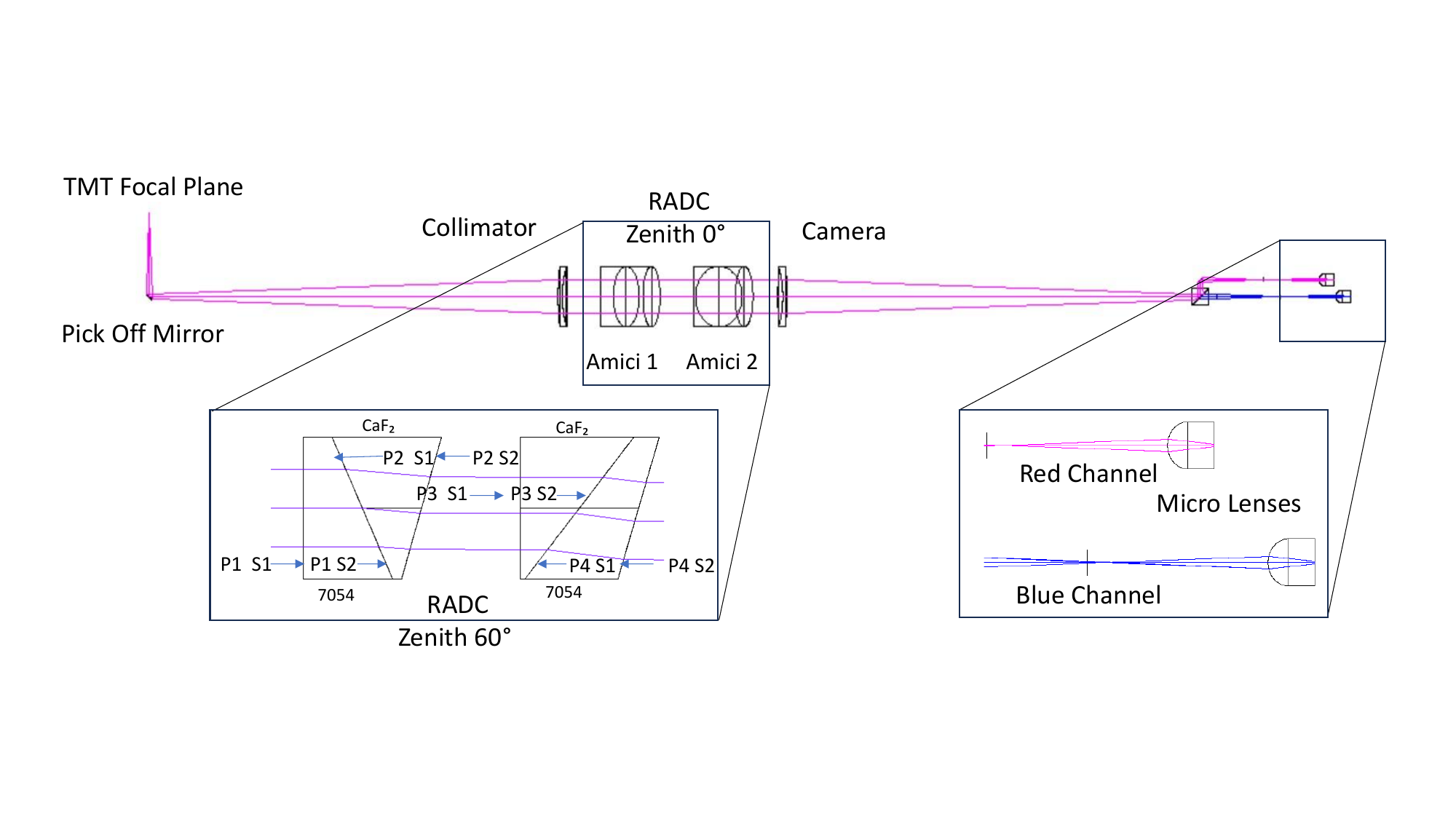} 
    \caption{Optical layout of the atmospheric dispersion corrector (ADC), where the Amici prisms are counter-rotated by 90 degrees at Zenith 0$^\circ$, along with the fore optics. P1, P2, P3, and P4 prisms belong to both Amici prisms. Here, $S$ denotes a surface, and S1 and S2 represent surface~1 and surface~2 of the prisms, respectively. In Amici 1, S1 and S2 correspond to P1, where S2 of P1 is S1 of P2, and S2 of P2 is the last surface of Amici 1. Similarly, in Amici 2, S2 of P3 is S1 of P4, and S2 of P4 is the last surface of Amici 2. The first inset shows the ADC at a Zenith angle of 60$^\circ$, incorporating flint glass Nikon-7054 and crown glass Ohara-CaF$_2$. The second inset displays the micro lens for the red and blue channels, which converts the f/15 beam to f/3.}

    \label{ADC}
\end{figure*}

\begin{table}
    \centering 
    \begin{tabular}{ll} 
        \hline
        \multicolumn{2}{c}{Specifications of TMT and HROS Instruments} \\
        \hline
        Primary mirror diameter & 30\,m \\
        Secondary mirror diameter & 3.1\,m \\
        Focal plane curvature radius & 3.01\,m \\
        Field of view size & 20\arcmin{} (equivalent to 2.616\,m) \\
        Plate scale  & $2.18\,\text{mm/1$\arcsec{}$}$ \\
        Wavelength range & 0.31--1\,$~\mu\text{m}$ \\
        Blue channel wavelength range & 0.31--0.45$~\mu$m\\
        Red channel wavelength range &  0.44--1.0$~\mu$m\\

        MOS spectral resolution & 25,000 \\
        Fiber size for MOS mode & 1\arcsec{} \\
        \hline
    \end{tabular}
    \caption{This table summarizes key specifications of the Thirty Meter Telescope (TMT) and the High-Resolution Optical Spectrograph (HROS).}
    \label{tmttable}
\end{table}

\section{ADC Design Considerations}

Achieving high radial velocity (RV) precision requires tight control over atmospheric dispersion residuals. To support an RV stability of 10 cm/s, residual dispersion must be maintained below 100 milliarcseconds (mas) across the 0.38–0.78 $\mu$m wavelength range \cite{Bachar_RV}. Assuming a linear relationship between RV precision and residuals, a stability target of 1 m/s allows residuals up to 1000 mas. Therefore, the required residual correction for our design is not difficult and does not impose severe constraints on system performance.

 To the date, ADCs compatible with the HROS working wavelength have not been reported, which spans from 0.31 to 1.0$~\mu\text{m}$. A reported three-material dispersion corrector using FK5, LLF2, and Fused Silica provides over 90\% transmission from 0.33--1.3$~\mu\text{m}$,  \citep{UV_ADC}. However, the material LLF2 has a transmission of approximately 20\% at 0.31$~\mu\text{m}$ for a 10~mm thickness\footnote{\url{https://refractiveindex.info/?shelf=specs&book=HIKARI-optical&page=LLF2}}, which is significantly lower than required, making it not suitable for our purposes. Therefore, designing the ADC for HROS-MOS requires the following:  
\begin{enumerate}
    \item Selection of an optimal material combination for the Amici prism to achieve a transmission efficiency greater than 90\% across the 0.31--1.0$~\mu\text{m}$ working wavelength range.
    \item Ensuring effective dispersion correction in the near ultraviolet region (below 0.33$~\mu\text{m}$).
    \item Maintaining dispersion corrected light within a 1\arcsec{} fiber diameter by controlling the beam's positional deviations as a function of zenith angle. This prevents introducing large movements of the fiber at the end of the fiber positioner \citep{Manju_Positioner}.
\end{enumerate} 

The limited availability of suitable materials and the need to correct for a higher dispersion towards the near UV region make designing an ADC for the HROS-MOS mode challenging. 

\label{design_considerations}

\section{Implementation of the Filippenko 1982 Model in Zemax}
We have used ZEMAX to design and analyze the optical system. The default atmospheric model in ZEMAX is based on the work of \cite{hohenkerk1985refraction}, accessed through the atmospheric surface feature option available in Zemax \citep{ZemaxManual}. However, \cite{ZEMAX_MODEL_COR} highlighted that this built-in model underestimates atmospheric dispersion, particularly in the blue spectral region. To mitigate this issue, we implemented the \cite{filippenko1982refraction} model, which has been extensively validated against on-site measurements of atmospheric dispersion \citep{bachar_dipersion_m}. This implementation was achieved by developing a Dynamic Link Library (DLL) file using C++ programming and incorporating it as a user-defined surface within Zemax. The source code and the DLL file are available in  \href{https://github.com/bestha95/Filippenko1982_DLL/}{GitHub}
\footnote{\url{https://github.com/bestha95/Filippenko1982_DLL/}}.

For the design of TMT-HROS, we adopted atmospheric parameters mentioned in \cite{Jason_KPIC_ADC} for the site Mauna Kea, at Hawaii, with a temperature of 276.15 $K$, relative humidity of $20\%$, and a pressure of 614 $mbar$. These parameters were then input into the newly implemented Filippenko model in Zemax to simulate the atmospheric conditions at the TMT-HROS site.

\label{filip_validation}

\section{Design of the Rotational ADC for HROS-TMT}

Although the elongation of spots due to dispersion for the entire wavelength range of HROS is $3.631\arcsec$, the dispersion for the blue and the red channels is $2.219\arcsec$ and $1.526\arcsec$, respectively. However, individually correcting the dispersion for both channels would require two separate correctors within the fiber positioner of the HROS-MOS mode. To avoid the need for two separate correctors that can be integrated within the positioners, we have designed a single corrector capable of working in the HROS passband.

First, we have modeled the Thirty Meter Telescope (TMT) with a primary mirror diameter of $30$ meters and a secondary mirror diameter of $3.1$ meters. The conic constants, representing hyperbolic mirrors, are $-1.0009530$ and $-1.3182280$, respectively. The radii of curvature are $-60000.0$ mm and $-6227.680$ mm. The distances between M1 and M2, and M2 and M3, are $-27093.750$ mm and $23593.750$ mm, respectively. A fold mirror (M3) diverts the beam to the Nasmyth platform at a distance of $19999.786$ mm, resulting in an effective focal length of 450 m and an f/15
\footnote{\url{https://www.tmt.org/}; \url{https://www.nao.ac.jp/en/research/telescope/tmt.html}}. The optical layout of the instrument and image quality of the TMT design are depicted in Figure~\ref{TMT_Design}. 

The design of ADCS includes a pick-off mirror of $\approx$ 5\arcsec{} diameter, which can tilt to direct the non-telecentric f/15 beam from the TMT focal plane, either on-axis or off-axis, to the doublet lens collimator, with an effective focal length of 291 mm, which collimates the dispersed beam, and then traverses through the ADC. As discussed in Section~\ref{Introduction}, an RADC is employed instead of an LADC to correct dispersion and deviations induced by the corrector through additional optimization to the prism surface tilts. The RADC comprises two doublet prisms, also known as Amici prisms, each consisting of two circular prisms (P1, P2, P3, P4, respectively) as shown in Figure \ref{ADC}.

The prisms were modeled in Zemax using tilted surfaces. A custom catalog was created to identify the optimal materials for the prisms, which consist of materials with transmission efficiencies ranging from $\approx$ $90\%$ to $100\%$ across a wavelength range of 0.31 to 1$~\mu\text{m}$. Materials are taken from Zemax inbuilt catalogs, such as Schott and Ohara glass catalogs, and a few materials are taken from \cite{nikon_i_line_e}. Table~\ref {material_table} provides the list of materials in the custom catalog. The transmission efficiencies of glass materials were calculated for a thickness of 10 mm as described in Equation~\ref{eq:beer_lambert}  derived from the Beer-Lambert law.

\begin{center}
\begin{minipage}{0.85\linewidth}
\begin{equation}
    T_{\text{n}} = T_{\text{o}}^{\frac{d_{\text{n}}}{d_{\text{o}}}},
    \label{eq:beer_lambert}
\end{equation}
\end{minipage}
\end{center}

where $T_{\text{n}}$ is the transmission for the new thickness, $T_{\text{o}}$ is the transmission for the original thickness (as provided in the vendor’s data sheet), $d_{\text{n}}$ and $d_{\text{o}}$ are the new and original material thicknesses, respectively.

\begin{table}
\centering
\renewcommand{\arraystretch}{0.5} 
\begin{tabular}{|l|l|l|l|}
\hline
Glass Name & Abbe Number & T: 0.31$~\mu\text{m}$ & T: 1$~\mu\text{m}$ \\ \hline
Nikon 7054         & 54.695634           & 0.9650                        & 0.9990                         \\ \hline
Nikon 5859         & 59.473084           & 0.9800                        & 0.9980                         \\
\hline
F\_SILICA          & 67.821433           & 1.0000                        & 1.0000                         \\ \hline
LITHOSIL-Q         & 67.826674           & 0.9996                        & 0.9997                         \\ 
\hline
S-FSL5             & 70.2363             & 0.8900                        & 0.9980                         \\ \hline
S-FSL5Y            & 70.354425           & 0.9540                        & 0.9980                         \\ \hline
N-FK5              & 70.405771           & 0.9545                        & 0.9990                         \\ \hline
FK5                & 70.405771           & 0.9502                        & 0.9987                         \\ \hline
S-FPL51Y           & 81.138165           & 0.8900                        & 0.9980                         \\ \hline
Nikon 4786         & 86.7798             & 0.9900                        & 0.9980                         \\ \hline
CaF$_2$            & 94.995854           & 1.0000                        & 1.0000                         \\ \hline
LITHOTEC-CAF2      & 95.232905           & 0.9992                        & 0.9996                         \\ \hline
NICF-V             & 95.260792           & 0.9980                        & 0.9980                         \\ \hline
\end{tabular}

\caption{Abbe number and internal transmittance values at 0.31~$\mu$m and 1~$\mu$m for the subset of glasses selected from the customized glass catalogue used during the optimization process. These materials were taken from Zemax inbuilt catalogs (such as Schott and Ohara) and a few additional materials from Nikon (2021). Only glasses with sufficiently high transmission were included. Transmission values are given as dimensionless fractions (per 10~mm thickness).}

\label{material_table}
\end{table}

The prism materials, apex angles, and counter-rotation angles of the Amici prisms were optimized using the \texttt{CENY}, \texttt{GLCX}, \texttt{GLCY}, and \texttt{DIFF} operands of Zemax in the merit function. Specifically, \texttt{CENY} tracks the Y-coordinate of the image centroid, while \texttt{GLCX} and \texttt{GLCY} assess the global X and Y positions of rays, and \texttt{DIFF} calculates the difference between operands, ensuring minimized deviations. This optimization process used the multi-configuration editor and Hammer optimization in Zemax.

\begin{figure}
    \centering
    \includegraphics[width=0.5\textwidth]{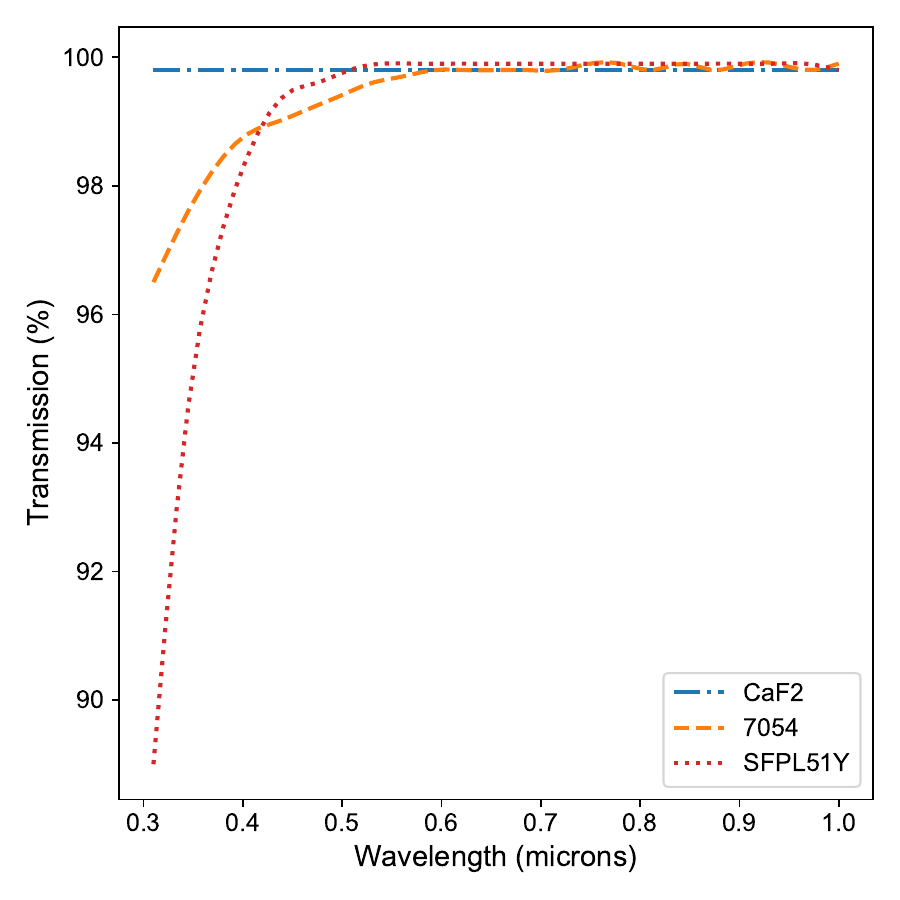}
    \caption{Transmission of materials used in designing the ADC (Nikon-7054 and CaF$_2$), as well as the collimator and camera (S-FPL51Y), for a thickness of 10 mm.}

    \label{transmission of materials}
\end{figure}
From the optimization, we identified that Nikon 7054 combined with CaF$_2$ from our custom catalog as the most suitable material combination for the prisms, exhibiting over 95\% transmission (per 10 mm thickness) over the 0.31--1.0$~\mu\text{m}$ range (See Figure \ref{transmission of materials}). The apex angles of the four individual prisms were determined such that the second prisms (P2, P4) corrected the deviation caused by the first prisms (P1, P3) within the Amici prisms. Consequently, (P1, P4) and (P2, P3) are using identical materials but oriented in reverse directions. The individual prisms have distinct surface tilts as mentioned below; the reason for the distinct surface tilts is explained in the section \ref{sec:deviations}.

\begin{itemize}
    \item P1: \( S_1 = 0^\circ, \, S_2 = -23.122^\circ \)
    \item P2: \( S_1 = -23.122^\circ, \, S_2 = 15.695^\circ \)
    \item P3: \( S_1 = 0^\circ, \, S_2 = 37.524^\circ \)
    \item P4: \( S_1 = 37.524^\circ, \, S_2 = 16.224^\circ \)
\end{itemize}

Here, \( P \) stands for Prism and \( S \) for Surface. The apex angle is calculated as $|S_2 - S_1|$, where $S$ denotes a surface, and $S_1$ and $S_2$ refer to tilt angles of surface 1 and surface 2 of the prisms, respectively. The two Amici prisms are then counter-rotated to cancel dispersion. The optimised parameters are provided in Table~\ref{adc_table}.

\begin{figure*} 
    \centering
    \includegraphics[trim={0cm 0cm 0.1cm 0cm},clip,width=\textwidth]{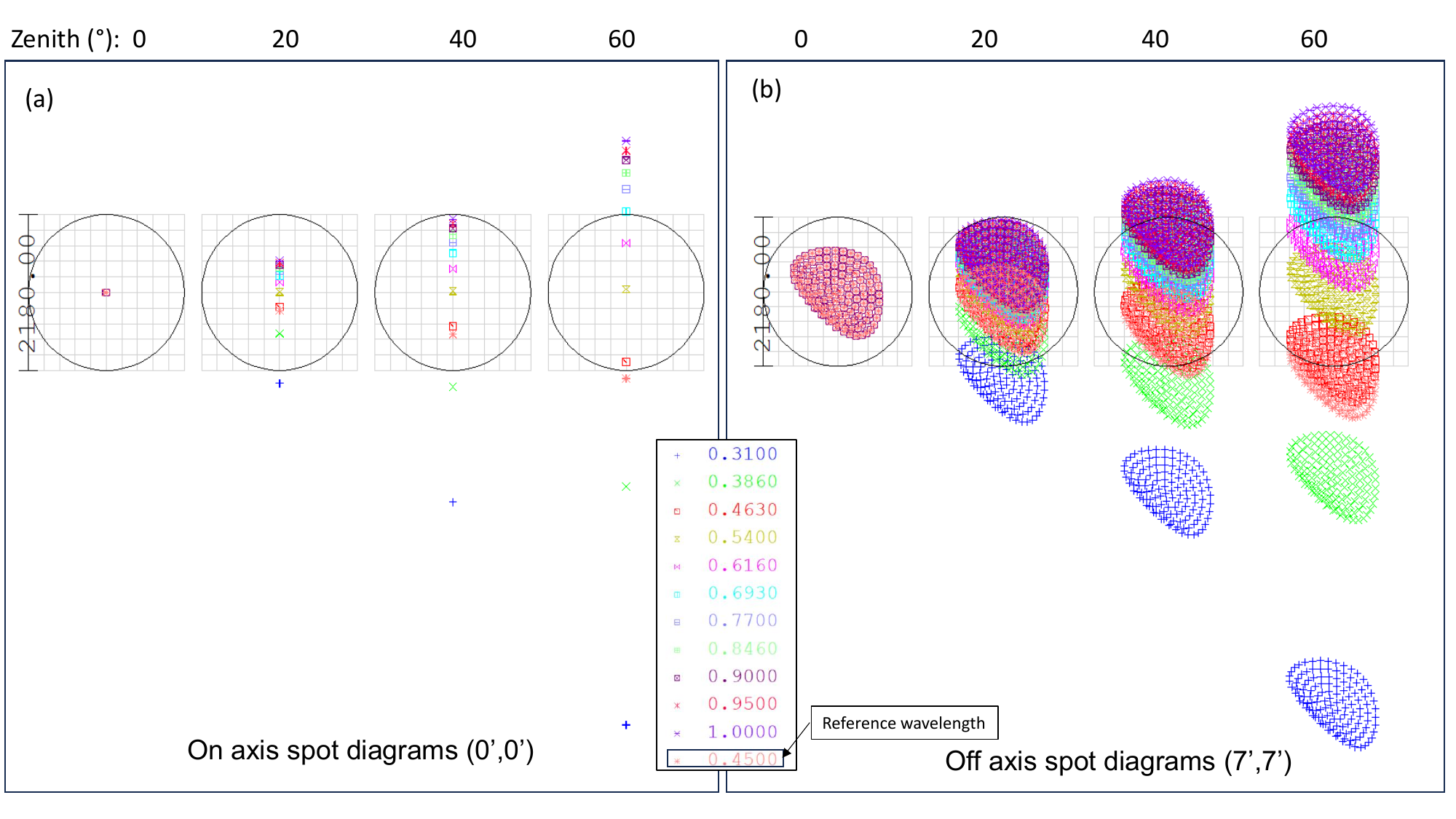}
    \caption{The image quality on the TMT focal plane. The circle represents a 1\arcsec{} fiber, corresponding to 2180 microns in linear scale at an f/15. (a) Spot diagrams of on-axis objects at different zenith angles without dispersion correction, where aberrations are minimal. (b) Spot diagrams of off-axis objects at different zenith angles without dispersion correction, where astigmatism predominantly affects image quality.}

    \label{uncorrected_spots_tmt_focal_plane}
\end{figure*}

\begin{figure*} 
    \centering
    \includegraphics[trim={0cm 6cm 0.1cm 2.5cm},clip,width=\textwidth]{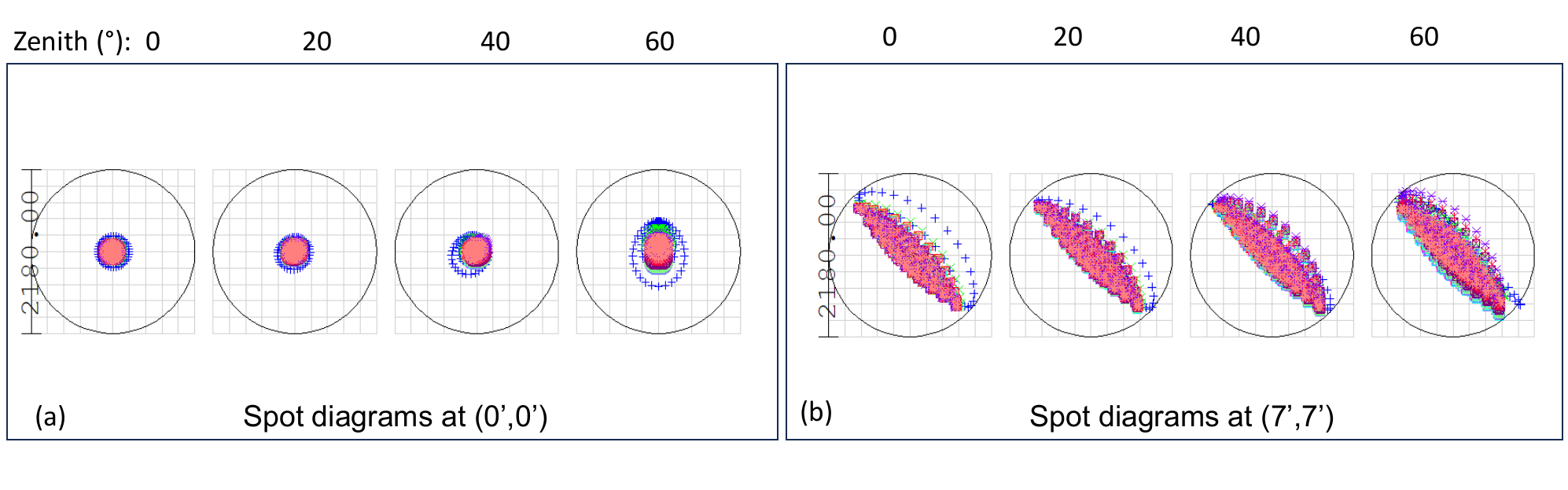}
    \caption{The image quality on the TMT focal plane. The circle represents a 1\arcsec{} fiber, corresponding to 2180 $~\mu\text{m}$ in linear scale at an f/15. (a) Spot diagrams of on-axis objects at different zenith angles after the dispersion correction (b) Spot diagrams of off-axis objects at different zenith angles after the dispersion correction}

    \label{corrected_spots_tmt_focal_plane}
\end{figure*}

Subsequently, a camera with specifications identical to the collimator focuses the dispersion-corrected beam exiting the ADC. The geometric spot diagrams before and after dispersion correction at the TMT focal plane are shown in Figures~\ref{uncorrected_spots_tmt_focal_plane} and~\ref{corrected_spots_tmt_focal_plane}, respectively. As shown in Figure~\ref{uncorrected_spots_tmt_focal_plane}(a), before dispersion correction, the on-axis spot sizes for each wavelength are very small, limited by the geometric aberrations of the RC design. After passing through the collimator and camera, the spots broaden due to spherical and chromatic aberrations, as shown in Figure~\ref{corrected_spots_tmt_focal_plane}(a). For the off-axis case, illustrated in Figure~\ref{uncorrected_spots_tmt_focal_plane}(b), the spots are dominated by astigmatism, and their sizes further increase due to the aberrations from the collimator and camera. Moreover, the tilted surfaces in the optical system alter the astigmatism predominantly \citep{Smith2023}, causing the spots to appear stretched (see Figure~\ref{corrected_spots_tmt_focal_plane}(b)).

\begin{figure*}
    \centering
    \includegraphics[width=1\textwidth,trim=0pt 0pt 0pt 0pt,clip]{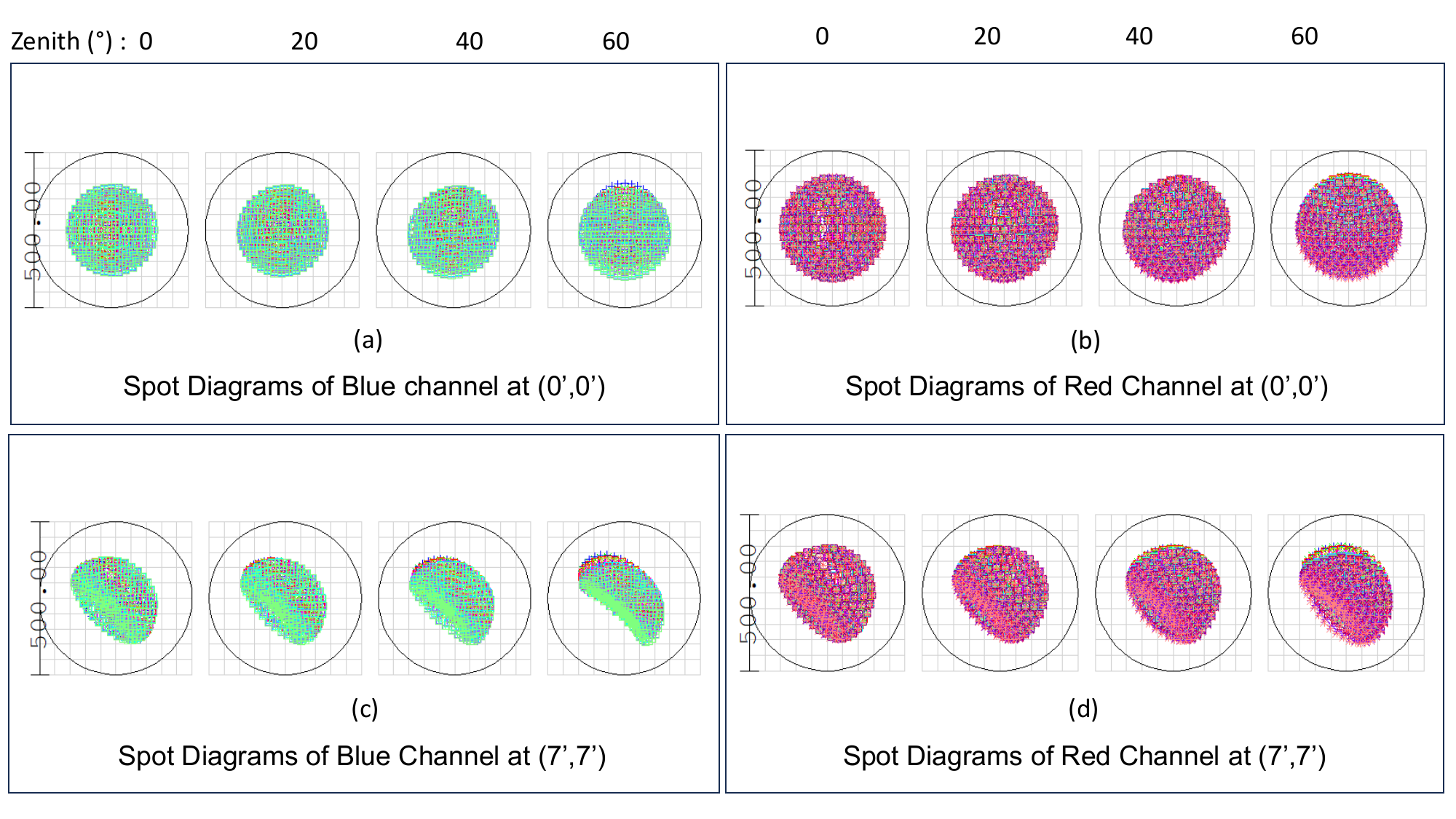}
\caption{The image quality of dispersion-corrected light after passing through the micro lens onto a 1" fiber, corresponding to 500$~\mu\text{m}$ in linear scale at an $\approx$ f/3 across different zenith angles. (a) Spot diagram of the on-axis object in the blue channel. (b) Spot diagram of the on-axis object in the red channel. (c) Spot diagram of the off-axis object in the blue channel. (d) Spot diagram of the off-axis object in the red channel.}
    \label{corrected_spots_after_adc}
\end{figure*}

Additionally, a dichroic\footnote{Since a dichroic with a cutoff wavelength near 0.45~$\mu$m is currently unavailable (\url{https://www.asahi-spectra.com/opticalfilters/astronomy_dichroic.asp}), a plane mirror at 45$^{\circ}$ embedded within a 15~mm silica cube is used as an alternative dichroic element. In this configuration, the tilted mirror is used to reflect light into the reflection channel, while the cube transmits the remaining light into the transmission channel.} is placed in the converging beam to separate the light into blue and red channels. The red channel is subsequently folded by a plane mirror to maintain the beam path within the fiber positioner. With an $\approx$ f/3 for both on-axis and off-axis objects across all zenith angles, the light is directed into both channels through microlenses made of Nikon~7054, positioned 33~mm from the camera focus.
The geometric spots of both channels on the fiber are shown in Figure \ref{corrected_spots_after_adc}. As shown, the image size at the pupil position (before the micro lens focus) increases from blue to red. This trend arises from the axial color of the micro lens, where the refractive index variation with wavelength causes shorter wavelengths to focus closer to the lens than longer wavelengths, resulting in better image quality for the blue channel and a progressively larger blur toward the red. For the off-axis case, the image at the pupil position appears perturbed since the micro lens was designed for the on-axis field. This distortion arises from the change in the solid angle (numerical aperture) of the off-axis fields, which alters the shape of the off-axis image at pupil position \citep{pupil_coma}. We adopted a slightly larger $500\,\mu$m fibre at an f/3 instead of the $436\,\mu$m fibre to simplify the calculation and illustration. Therefore, at an f/3, the plate scale is $0.5\,\mathrm{mm}$ per $1^{\prime\prime}$.

\begin{table}
    \centering

    \begin{tabular}{ll} 
        \hline
        \multicolumn{2}{c}{Atmospheric Dispersion Corrector} \\ [0.5ex]
        \hline
        Diameter & 30\,mm \\
        Materials & Nikon 7054 \& CaF$_2$ \\
        Apex Angles of P1, P2, P3, and P4 & 23.122 $^\circ$, 38.807$^\circ$, 37.524$^\circ$, 21.3$^\circ$ \\
        Counter rotation angles of Amici-prisms \\
        at zenith angles 0$^\circ$, 20$^\circ$, 40$^\circ$, 60$^\circ$ & 90$^\circ$, 78.331$^\circ$, 61.986$^\circ$, 1.672$^\circ$ \\
        \hline
        \multicolumn{2}{c}{Collimator and Camera} \\ [0.5ex]
        \hline
        Diameter & 30\,mm \\
        Effective focal length & 291\,mm \\
        Materials & S-FPL51Y \& Nikon 7054 \\
        \hline

        \hline
        \multicolumn{2}{c}{Micro Lens} \\ [0.5ex]
        \hline
        Semi Diameter & 3.6 \,mm \\
        Radius of Curvature & 3.6 \,mm \\
        Materials & Nikon 7054 \\
        \hline
    \end{tabular}
    \caption{Specifications of the atmospheric dispersion corrector, collimator, camera, and micro lens.}
    \label{adc_table}
\end{table}

\label{section5}
\section{Analysis}

Upon completing this design, we sought to validate whether it met our requirements. We performed dispersion residual calculations, beam deviation analyses, throughput calculations, and tolerance analyses to validate this design, which are detailed in this section.

\subsection{Residuals}

As discussed in section~\ref{design_considerations}, the residuals relevant to our design requirements are minimal. Nevertheless, we aimed to demonstrate the residual dispersion present in our design after correction. To achieve this, we utilized the ray-trace option in Zemax to compute the dispersion by converting the linear position of the rays at the TMT focal plane, as a function of zenith angle and wavelength, into an angular scale using the plate scale.

\begin{figure}
    \centering
    \begin{subfigure}{0.5\textwidth}
        \centering
        \includegraphics[width=\textwidth]{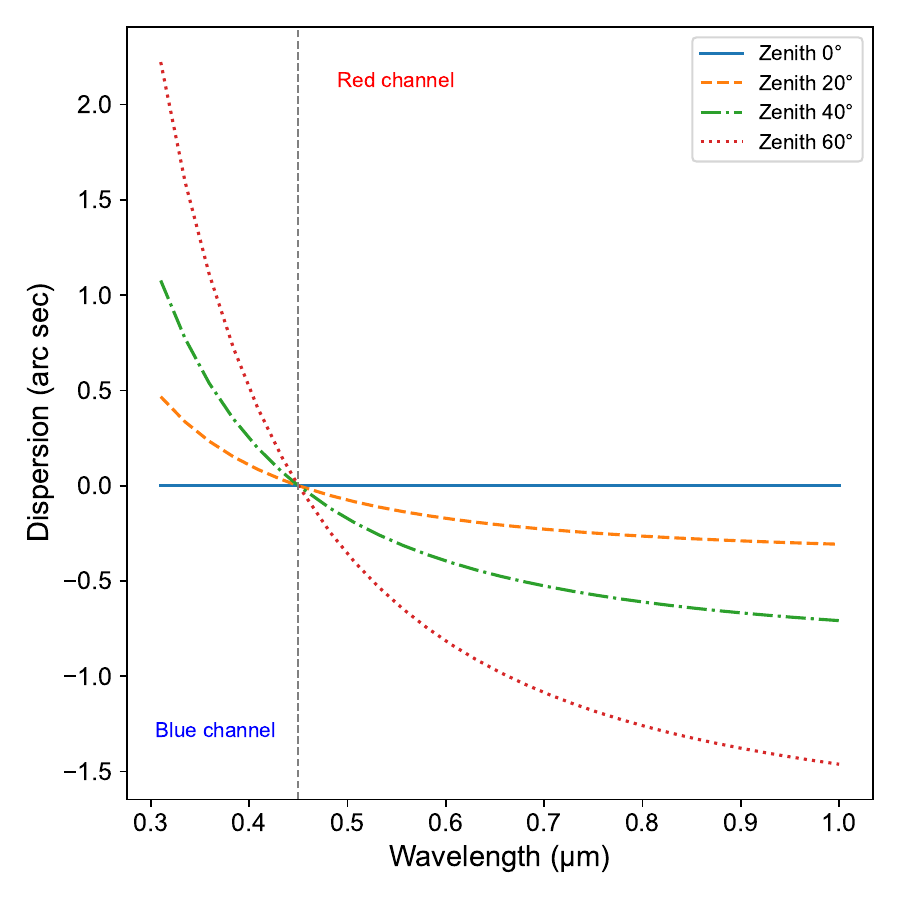}
        \caption{The dispersion at the telescope focal plane computed before correction at different zenith angles}
        \label{dispersion_before}
    \end{subfigure}
    \hfill
    \begin{subfigure}{0.5\textwidth}
        \centering
        \includegraphics[width=\textwidth]{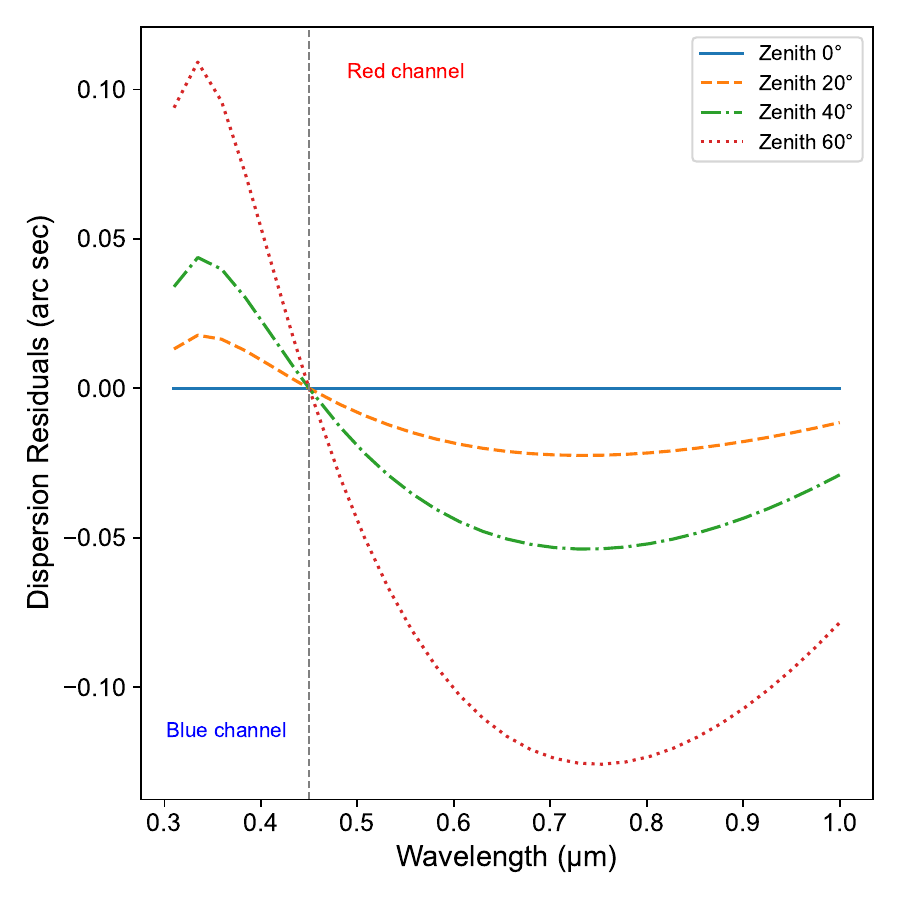}
        \caption{The dispersion residuals after correction computed before the micro lens at different zenith angles by replacing the collimator and camera with paraxial lenses. (Residuals)}
        \label{dispersion_after}
    \end{subfigure}
    \caption{The figure depicts the dispersion before and after correction across the HROS working wavelength range at different zenith angles, modeled using the Filippenko 1982 framework in Zemax.}

    \label{FilippenkoDispersion}
\end{figure}

To validate the performance of the Filippenko 1982 model (DLL file) in Zemax, we compared its results with those from a Python-based implementation of the same model \citep{bachar_dipersion_m} under identical atmospheric conditions. The difference was 0.1 mas, which is negligible as it is very small compared to the maximum dispersion at zenith angle 60$^\circ$ of HROS working wavelength coverage ($\approx$ 3.631\arcsec{}) as shown in Figure \ref{dispersion_before}, and well within the HROS requirements.

Subsequently, the collimator and camera were replaced with paraxial lenses to eliminate the dispersion contributions from the transmission elements, ensuring that only the ADC's residual dispersion was computed after correction. This residual dispersion was computed before the micro lens at an f/15 as a function of zenith angle, wavelength, and counter-rotation angle of the Amici prisms. These residuals are within 250 mas across the HROS working wavelength range, as depicted in Figure \ref{dispersion_after}. As shown, the residual dispersion after correction exhibits a nonlinear shape, due to the anomalous dispersion characteristics of the ADC glass materials \citep{anm_dispersion1}. This arises because the net dispersion curve of the corrector does not perfectly match the atmospheric dispersion over the wavelength range \citep{anm_dispersion2}. Nonetheless, these residuals remain within the requirements of HROS-ADC.

\subsection{Deviations}
\label{sec:deviations}
\begin{figure}
    \centering
    \includegraphics[width=0.5\textwidth,trim=0 1 0 0,clip]{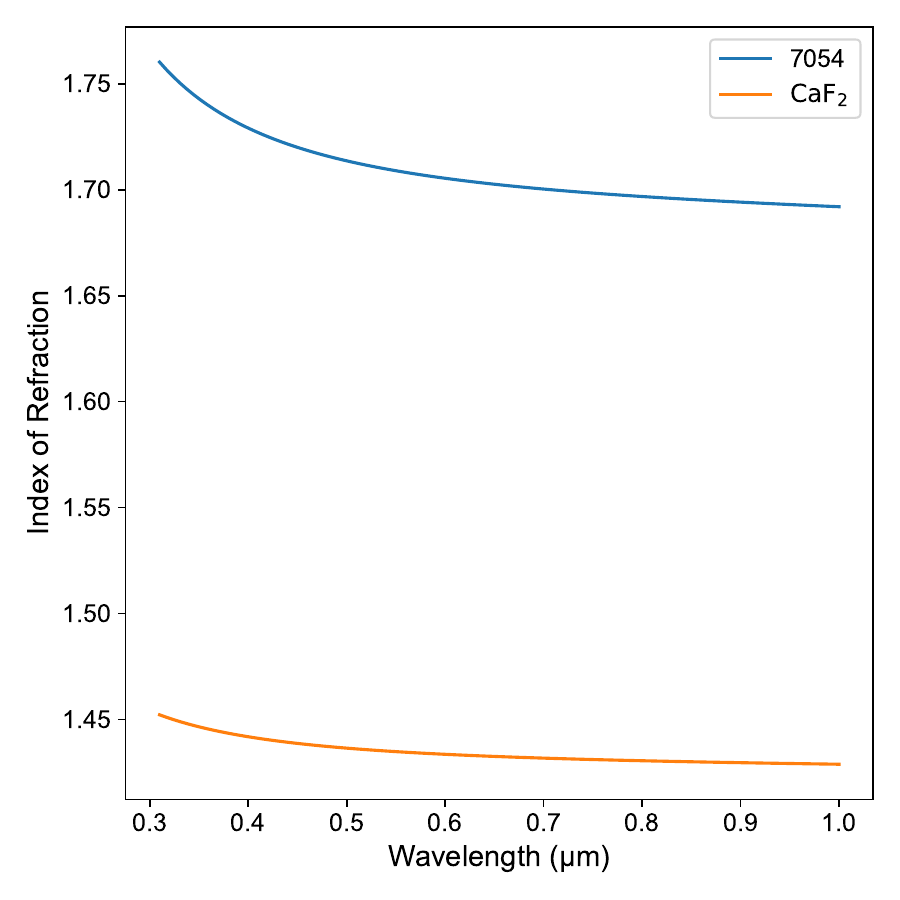}
    \caption{The dispersion curves of materials apart from each other, which are used in designing the ADC (Nikon-7054 and CaF$_2$)}

    \label{dispersion_curves}
\end{figure}

In principle, LADCs work by altering the incident angle of rays on the movable prism as a function of the zenith angle to correct atmospheric dispersion. Conversely,  RADCs function by changing the incident angle of rays relative to the prism normal, depending on the rotation of the Amici prisms. In LADCs, the beam deviations introduced by the first prism are offset by the second prism. In RADCs, a strategy involves using two different types of glass for the prisms, which exhibit varying dispersions but share the same refractive index at a specific wavelength \citep{Bahrami_11}. Initially, we designed the ADC using identical Amici prisms, which corrected the atmospheric dispersion, resulting in residuals of 80 mas over the 0.31–1.0$~\mu$m range, demonstrating good performance. However, After correction, the polychromatic beam is deviated by up to 17\arcsec{} at a zenith angle of 60\degree, which corresponds to 38~mm in linear scale at the collimator focal plane for an f/15 beam. These deviations are caused by atmospheric refraction and the ADC itself, and they vary with both zenith angle and prism counter-rotation angles. The large deviation at higher zenith angles arises because the dispersion curves of the selected materials differ significantly, as shown in Figure \ref{dispersion_curves}.

\begin{figure}
    \centering
    \includegraphics[width=0.5\textwidth,trim=0 10 0 6,clip]{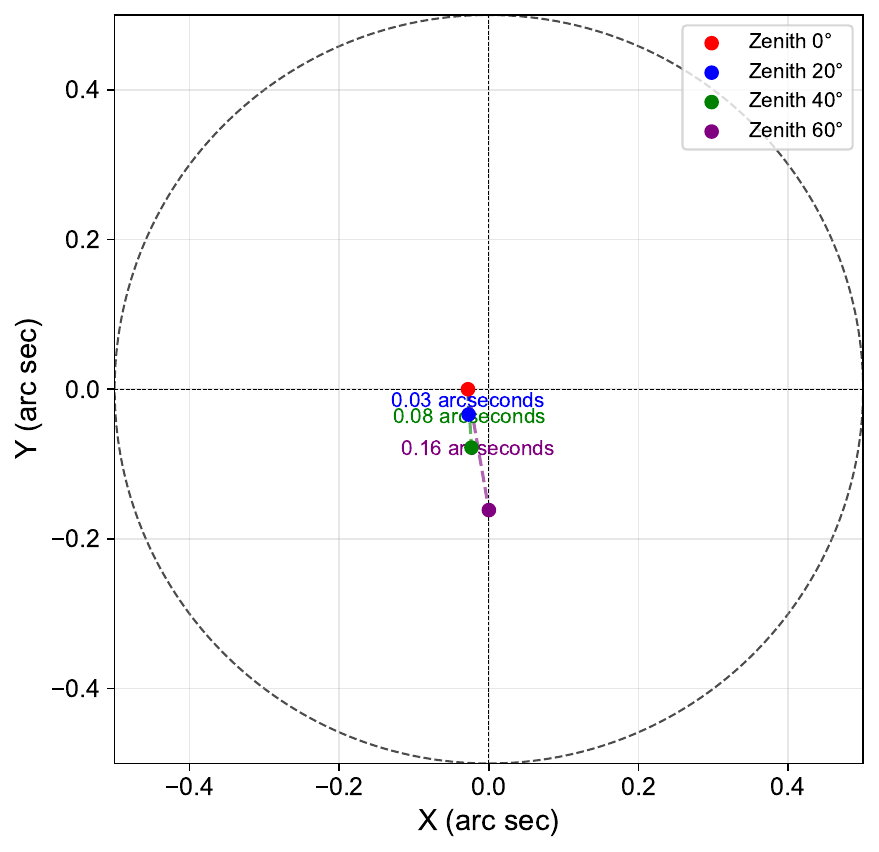}
    \caption{The figure shows a 1\arcsec{} fiber represented by a circle, with dots in different colors corresponding to the position of a reference wavelength (0.45$~\mu\text{m}$) at various zenith angles, as labeled in the image. The movement of the dots within the fiber illustrates the effect of deviation after the dispersion correction. It also demonstrates that the maximum deviation of the wavelength at 0.45$~\mu\text{m}$ remains within the 1\arcsec{} fiber after the dispersion has been corrected.}

    \label{Deviations}
\end{figure}

This substantial deviation was assessed using the ray tracing option in Zemax. The methodology involved calculating the distance between the position of the reference wavelength (0.45$~\mu\text{m}$) at zenith 0$^\circ$ and its positions at other zenith angles on the focal plane after dispersion correction. In this case, achieving maximum throughput requires displacing the fiber along the beam deviation direction, which closely coincides with the dispersion (parallactic) direction. However, this is challenging due to the significant beam deviation. To address this issue and avoid displacement of the fiber, tilts of outer surfaces of the Amici prisms were optimized, then the deviations were reduced to 0.16\arcsec{}, ensuring the beam remains well within the 1\arcsec{} of the fiber as shown in Figure \ref{Deviations}. Consequently, the displacements can be regarded as negligible.

\subsection{Throughput}

The geometrical image analysis (IMA) tool in \texttt{ZEMAX} and the \texttt{PyZDDE} Application Program Interface (API) \citep{indranil_2014} were used to estimate the geometric throughput for both on-axis and off-axis objects within a 1\arcsec{} fiber. The analysis was performed using image size settings of 15 mm and 3 mm based on the f-ratio of the beam. This analysis was conducted for the entire HROS working wavelength range, with and without dispersion correction. The methodology for calculating geometrical throughput involved calculating the seeing for each wavelength using the relationship among seeing, Fried's parameter, and wavelength, as described in Equation \ref{eq:fried}. We considered a seeing of 0.7\arcsec{} at a reference wavelength ($\lambda_{\text{ref}} = 0.45~\mu\text{m}$) and computed seeing at different wavelengths. Using that, we created 2D Gaussian profiles, which were then convolved using \texttt{fftconvolve} from the SciPy library \cite{2020SciPy-NMeth} to perform fast convolution with the extracted geometrical position-based intensities obtained from the IMA tool. This process generated images of seeing-included geometric spots for different zenith angles before and after dispersion correction (see Figure \ref{plots_with_seeing}). The final results were multiplied by a 1\arcsec{} python-simulated circular aperture to account for the fiber size.
The Fried parameter is given by ;

\begin{equation}
    r_0 \, (\text{m}) = \frac{0.98 \times \lambda\,(\text{m}) \times 206265}{\text{FWHM}_{\text{seeing}} \times \text{plate scale (m/arcsec)}},
    \label{eq:fried}
\end{equation}

where \( r_0 \) is the Fried parameter, representing the atmospheric coherence length. The constant 0.98 is empirical, 0.45$~\mu\text{m}$ (\(0.45\times 10^{-6} \) m) is the reference wavelength, and 206265 is a factor for converting radians to arcseconds. The term \( \text{FWHM}_{\text{seeing}} \) represents the full width at half maximum of the seeing disk at the reference wavelength \citep{frieds_parameter}.\\

\begin{figure}
    \centering
    \begin{subfigure}{0.5\textwidth}
        \centering
        \includegraphics[width=\textwidth]{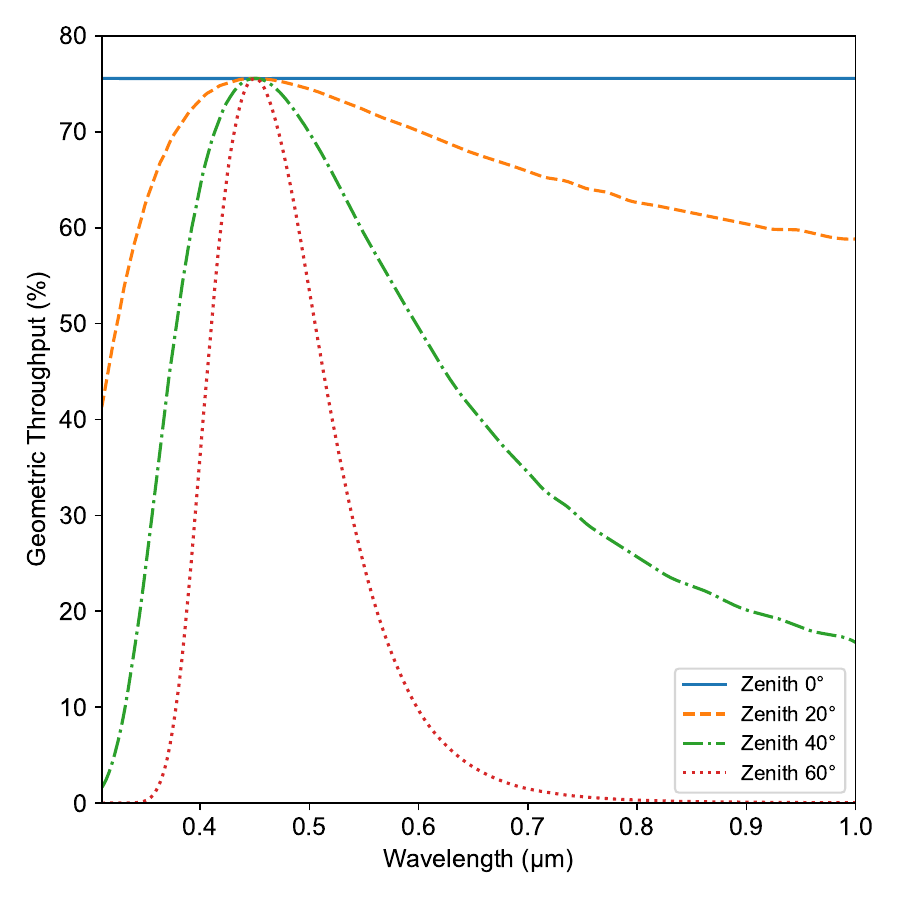}
        \caption{\hspace{10pt} Before the dispersion correction}
        \label{fig:onFilippenkoThroughputBefore}
    \end{subfigure}
    \hfill
    \begin{subfigure}{0.5\textwidth}
        \centering
        \includegraphics[width=\textwidth]{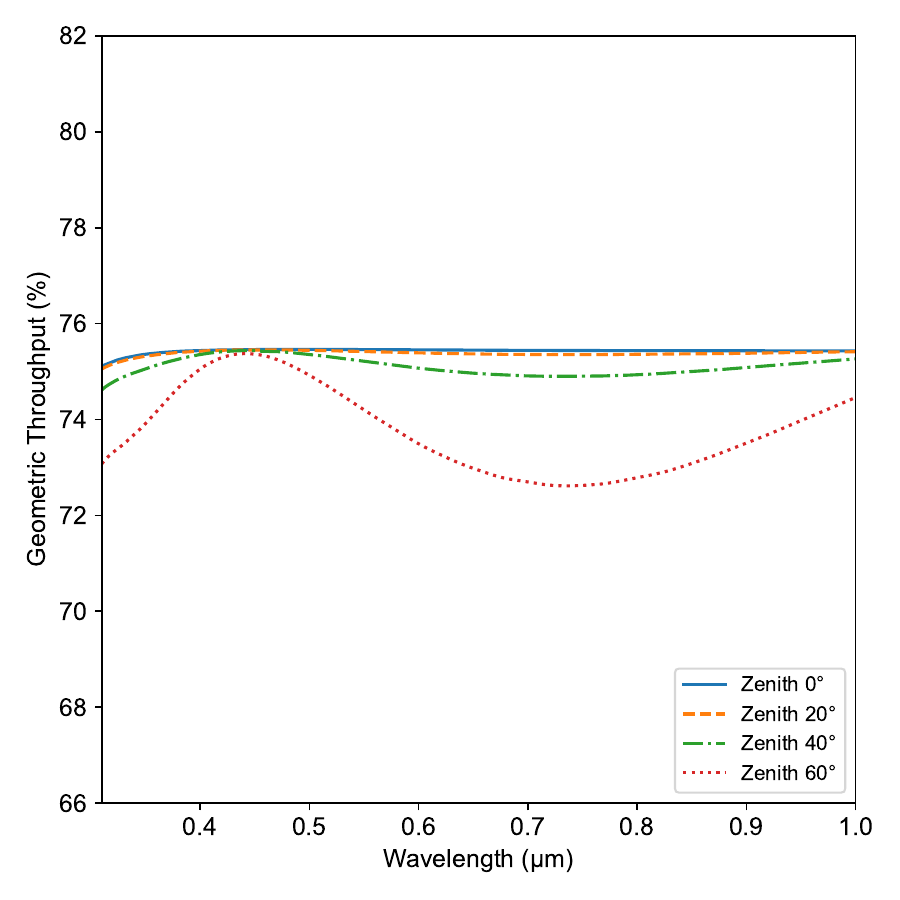}
        \caption{\hspace{15pt} After the dispersion correction}
        \label{fig:onFilippenkoThroughputAfter}
    \end{subfigure}
    \caption{The geometric throughput across the HROS working wavelength range for the on-axis object at different zenith angles, estimated before splitting into channels.}

    \label{fig:Filippenkothroughput_onaxis}
\end{figure}

\begin{figure}
    \centering
    \begin{subfigure}{0.5\textwidth}
        \centering
        \includegraphics[width=\textwidth]{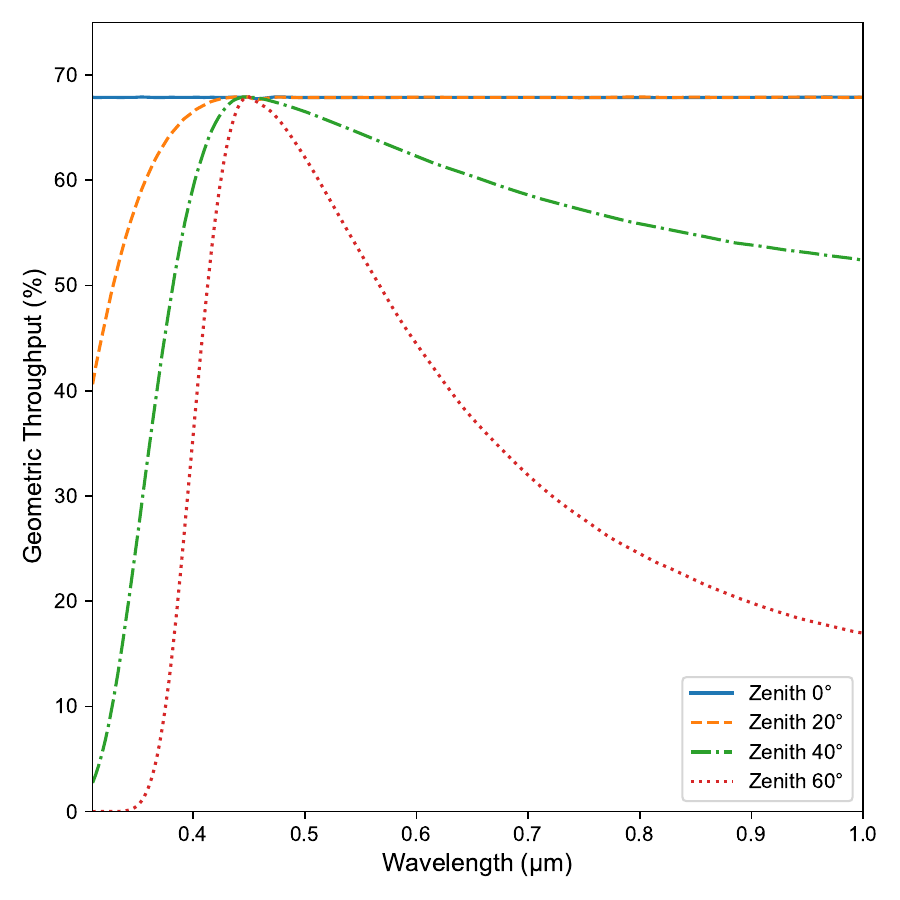}
        \caption{\hspace{10pt} Before the dispersion correction}
        \label{fig:FilippenkoThroughputBefore_field}
    \end{subfigure}
    \hfill
    \begin{subfigure}{0.5\textwidth}
        \centering
        \includegraphics[width=\textwidth]{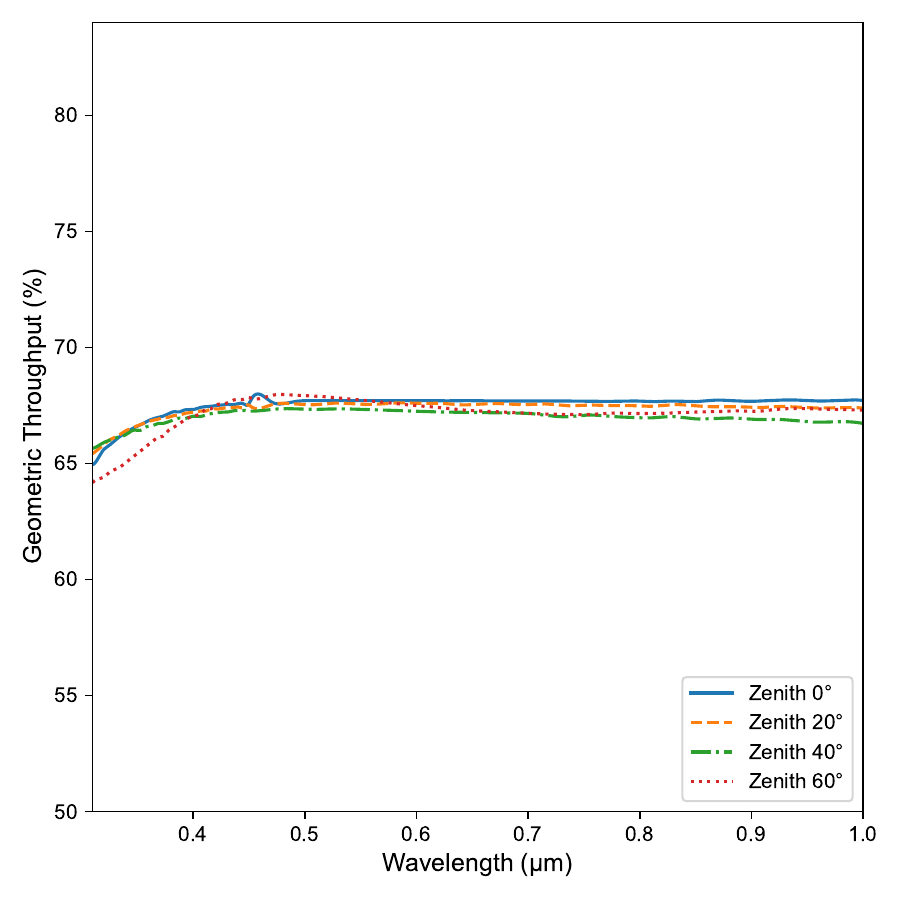}
        \caption{\hspace{15pt} After the dispersion correction}
        \label{fig:FilippenkoThroughputAfter_filed}
    \end{subfigure}
    \caption{The geometric throughput across the HROS working wavelength range for the off-axis object at different zenith angles, estimated before splitting into channels.}

    \label{fig:Filippenkothroughput_offaxis}
\end{figure}

The 2D Gaussian function is given by;

\begin{equation}
    \text{I (x,y)} = \frac{\exp\left(-\left(\frac{(x-x_0)^2}{2\sigma_x^2} + \frac{(y-y_0)^2}{2\sigma_y^2}\right)\right)}{2\pi\sigma_x\sigma_y},
    \label{eq:gaussian}
\end{equation}

where \( I \) is a function of displacement \((x, y)\) relative to the origin \((x_0, y_0)\). The parameters \( \sigma_x \) and \( \sigma_y \) represent the standard deviations of the distribution in arcseconds. In Figure \ref{plots_with_seeing}, the panel (a) illustrates the decrease of intensity along the dispersion direction on the blue side (high dispersion) and its increase on the red side (low dispersion), while panel (b) shows the same images after correcting the dispersion within the 1\arcsec{} fiber before entering the micro lenses. The percentage of geometric throughput ($\%GT$) is computed using the following equation:

\begin{figure*}
    \centering
    \includegraphics[width=1.05\textwidth,trim=50pt 70pt 0pt 50pt,clip]{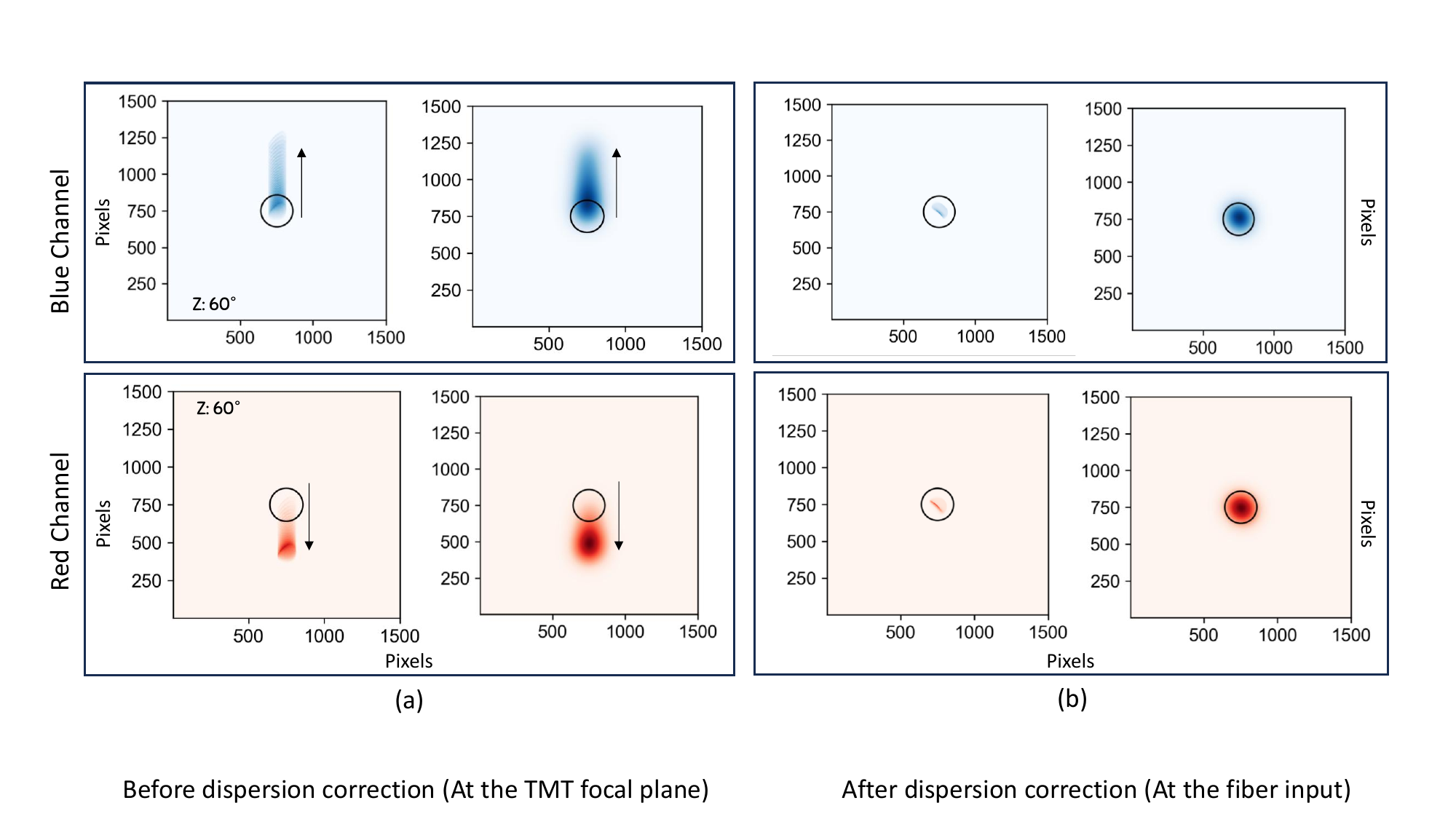}
    \caption{The figure illustrates aberrated off-axis spots at the reference wavelength (0.45$~\mu\text{m}$), positioned at the center. The blue and red colors represent the blue and red channels of the wavelength coverage, respectively. The scale of the above image is 218~pixels per $1^{\prime\prime}$. At an f/15 beam, 1\arcsec{} = 2.18 mm, while at an f/3 beam it corresponds to 0.5~mm. The black circle and arrow represent the $1^{\prime\prime}$ fibre and the direction of atmospheric dispersion, respectively. (a) Before the dispersion correction, the first column displays spot diagrams without atmospheric seeing, while the second column presents seeing-included images at the telescope focal plane. (b) After the dispersion correction, the first column shows spot diagrams without atmospheric seeing, whereas the second column illustrates seeing-included images at the fiber.}
    
    \label{plots_with_seeing}
\end{figure*}

\begin{gather}
    \% \text{GT} = \left( \frac{\text{OPI}}{\text{IPI}} \right) \times 100,
    \label{GP}
\end{gather}

where IPI (Input Intensity) represents the sum of intensity before multiplication by the 1\arcsec{} circular aperture, while OPI (Output Intensity) represents the sum of intensity after multiplication by the 1\arcsec{} circular aperture.\\
\noindent Using this methodology, percentages of geometrical throughput for different zenith angles for an on-axis object, both before and after dispersion correction, are obtained. However, aberrations such as astigmatism affect the off-axis throughput, which is more prominent in RC telescopes and leads to signal loss at the fiber. To estimate its effect on \% \text{GT}, the instrument's position was adjusted to analyze image quality over a (7\arcmin{},7\arcmin{}) field of view.
To account for the curvature of the TMT focal plane (sag = 279.214~mm over a (7\arcmin{},7\arcmin{}) field), the ADCS at the off-axis position was repositioned to match the focal coordinates \((x, y, z) = (912, 912, -279.214)\)~mm, corresponding to a distance of 19720.786~mm from M3 along the optical axis and 912~mm along each of the two perpendicular axes, ensuring that the off-axis focal surface coincides with the ADCS during throughput evaluation \citep{Manju_Positioner}.
The off-axis images were then evaluated, including the impact of atmospheric seeing. The results of \% \text{GT} from these analyses are presented in Figure~\ref{fig:Filippenkothroughput_onaxis} and \ref{fig:Filippenkothroughput_offaxis}. The plots show a significant increase in throughput after dispersion corrections. At a Zenith angle of 0$^\circ$, the on-axis \% \text{GT} is $\approx 75\%$ and off-axis \% \text{GT} is $\approx 68\%$ at 0.31$\mu$m. At a Zenith angle of 60$^\circ$, both on-axis and off-axis \% \text{GT} were initially 0\% at 0.31$\mu$m, but after correction, on-axis \% \text{GT} increased to 73\% and off-axis \% \text{GT} to 65\%.

As shown in Figures \ref{fig:Filippenkothroughput_onaxis} and \ref{fig:Filippenkothroughput_offaxis}, the maximum throughput occurs at the reference wavelength, as it lies closest to the optical axis after dispersion correction, and this is computed before the micro lens. These calculations assume the fiber is positioned at the reference wavelength location, ensuring coverage across the entire HROS working
wavelength range. However, if the fibers in both channels are offset relative to their current positions, the maximum throughput of the wavelength will shift accordingly. Also, due to the non-linear nature of the dispersion correction, dips can be seen on the bluer side and around 0.72~$\mu$m in the on-axis plot (see Figure \ref{fig:onFilippenkoThroughputAfter}), where the corrected dispersion dominates over the optical aberrations. However, a similar trend is not observed in the off-axis plot (see Figure \ref{fig:FilippenkoThroughputAfter_filed}), where the optical aberrations dominate over the corrected dispersion.

To calculate the input signal entering the spectrograph from a 1\arcsec{} fiber, we considered factors, including the system's polishing quality of reflective elements ($pq$) which is the product of reflectivity of M1, M2, M3, and the pick-off mirror, Fresnel input-output losses ($f_l$), lens misalignment errors ($m_e$) and focal ratio degradation ($FRD$). The typical values assumed are \( pq = 0.92 \), $f_l$ = $0.92$, $m_e$ = $0.98$ and $FRD$ = $0.90$.  These values account for simple spectrograph design throughput estimation as given in \cite{Avila2017}. The transmission of high OH fiber ($t_f$) over 20 meters was calculated using the attenuation as a function of wavelengths provided by \href{https://www.thorlabs.com/Images/TabImages/FG_UGA_LGA_UCA_LCA_UEA_LEA_Attenuation_Data_2.xlsx}{Thorlabs}
\footnote{\url{https://www.thorlabs.com/Images/TabImages/FG_UGA_LGA_UCA_LCA_UEA_LEA_Attenuation_Data_2.xlsx}} using Equation~\ref{eq:fiber_transmission}. Furthermore, the transmission of the lenses (10 mm thickness) and prisms (50 mm thickness) within the system as a function of wavelength was also considered.
The final calculation of the fiber output is illustrated in Equation~\ref{eq:spectrograph_input}.

\begin{equation}
\begin{aligned}
    T_{\text{km}} &= 10^{-\alpha / 10}, \\
    T_{\text{fiber}} &= T_{\text{km}}^{L / 1000},
\end{aligned}
\label{eq:fiber_transmission}
\end{equation}

where \( \alpha \) is the fiber attenuation in dB/km, \( L \) is the fiber length in meters, \( T_{\text{km}} \) is the transmission for 1 km of fiber, and \( T_{\text{fiber}} \) is the total transmission through a fiber of length \( L \).

\begin{align}
    I_{\text{in}}(\lambda, Z) &= pq \cdot f_l \cdot m_e \cdot FRD \cdot f_t(\lambda) \cdot t_{fa}(\lambda) \cdot \text{\%GT}(\lambda, Z),
    \label{eq:spectrograph_input}
\end{align}

In this expression, \( I_{\text{in}}(\lambda, Z) \) is the input efficiency to the spectrograph, dependent on wavelength \( \lambda \) and zenith angle \( Z \). The term \(t_f(\lambda) \) denotes the fiber transmission, \( t_{fa}(\lambda) \) transmission of fore optics (excluding dichroic and micro lens) and ADC, and \( \text{\%GT}(\lambda, Z) \) represents the geometric throughput.

\begin{figure}
    \centering
    \begin{subfigure}{0.5\textwidth}
        \centering
        \includegraphics[width=\textwidth]{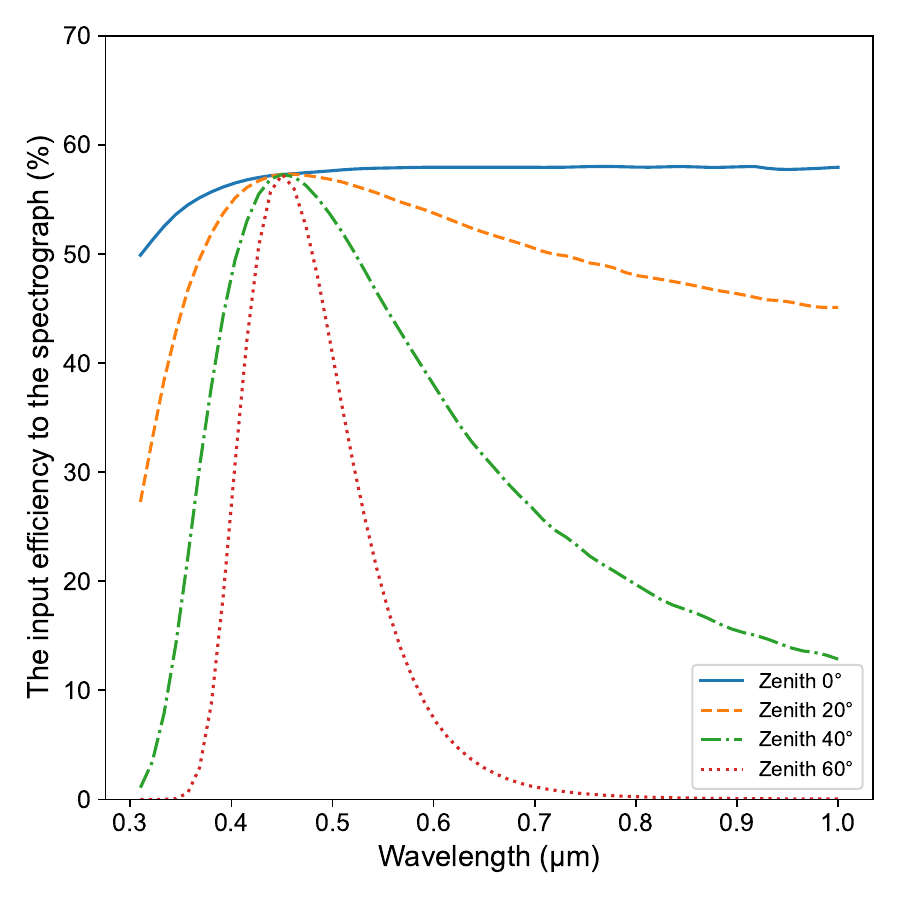}
        \caption{\hspace{15pt}Before the dispersion correction}
        \label{input_efficiency_before}
    \end{subfigure}
    \hfill
    \begin{subfigure}{0.5\textwidth}
        \centering
        \includegraphics[width=\textwidth]{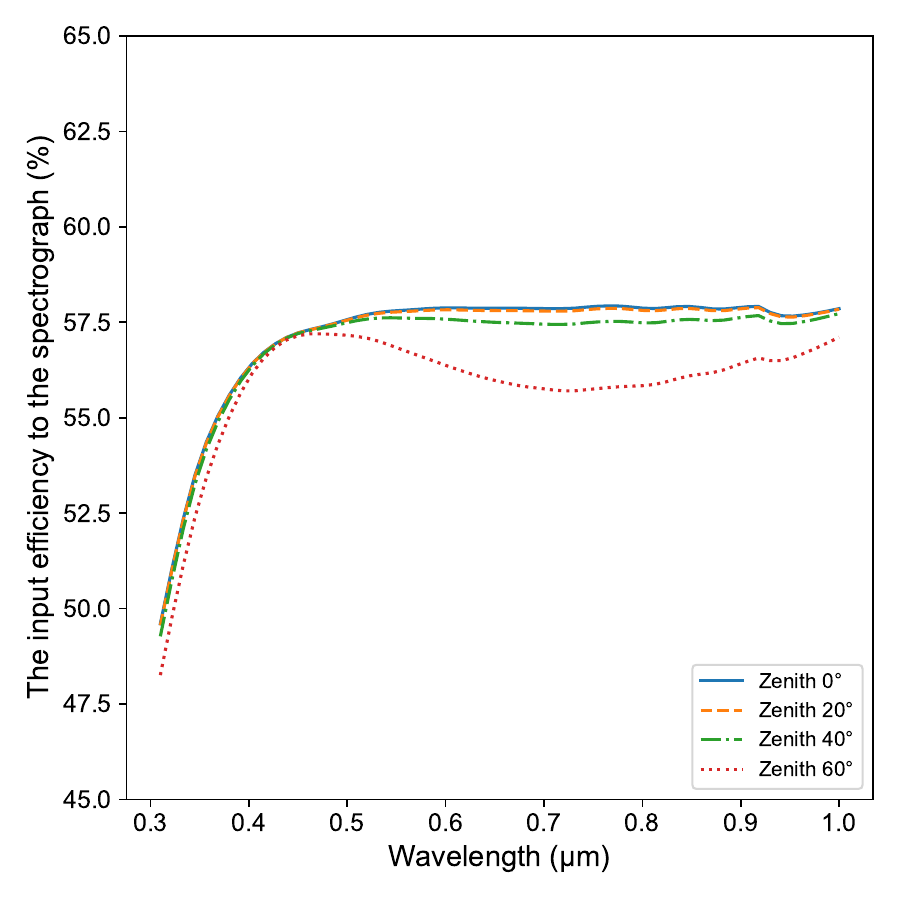}
        \caption{\hspace{15pt} After the dispersion correction}
        \label{input_efficiency_after}
    \end{subfigure}
    \caption{The total throughput entering the spectrograph across the HROS working wavelength range, as estimated by accounting for the transmission efficiency of the dichroic and micro lens.}

    \label{Efficiency}
\end{figure}

\begin{figure}
    \centering
    \includegraphics[width=0.5\textwidth]{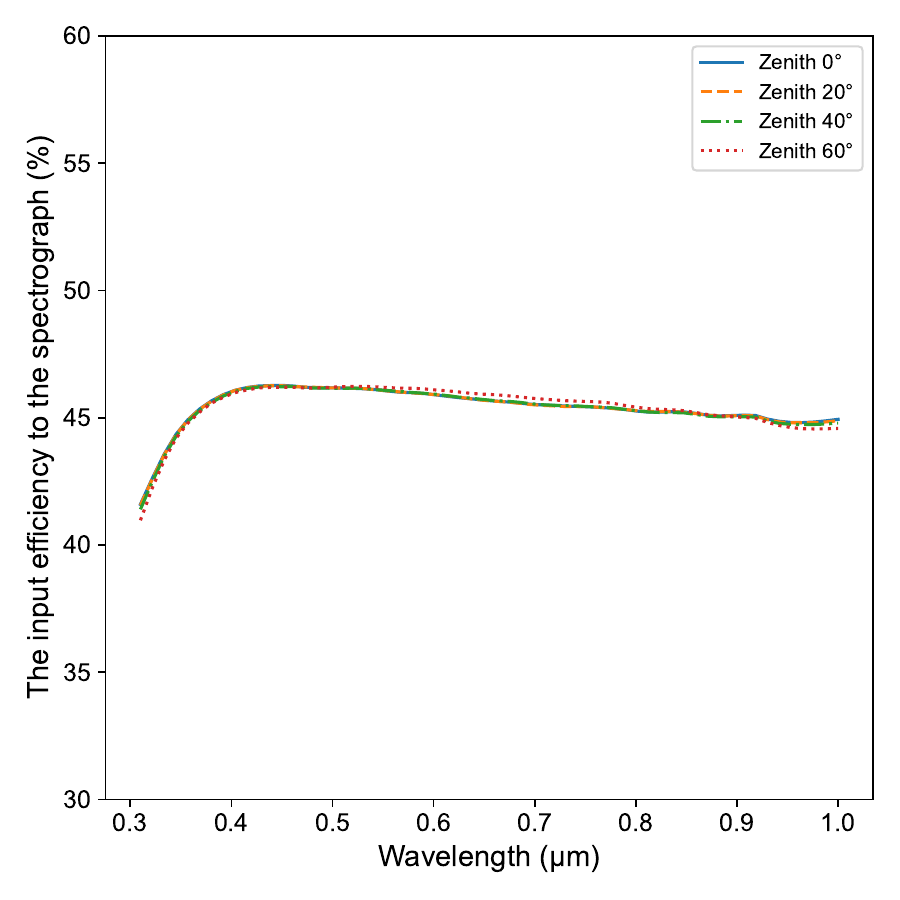}
    \caption{The total throughput entering the spectrograph across the HROS working wavelength range, as estimated by considering the optical effects of the micro lens.}

    \label{Efficiency_inluding_microlense}
\end{figure}

\noindent The input efficiency to the spectrograph, as a function of wavelengths and zenith angles, has significantly increased for the on-axis object after dispersion correction. At Zenith 60$^\circ$, the efficiency was zero before the correction but improved to above 45\% after the correction. The results are presented in Figures~\ref{input_efficiency_before} and~\ref{input_efficiency_after}.

In the above efficiency calculations, the optical performance of the micro lens and dichroic was not considered, as the influence of atmospheric dispersion on throughput is not seen clearly at its pupil position, where aberrations such as spherical and axial chromatic aberration are present. The image at pupil position is considered to ensure uniform illumination across the fiber. The input efficiency to the spectrograph was estimated using the percentage geometric throughput (\%GT) at the pupil position, incorporating a dichroic transmission of 90\% and the micro lens transmission across the full wavelength range, corresponding to lengths of 15~mm and 10~mm, respectively, along with the other parameters used in the previous efficiency calculation.
The input efficiency to the spectrograph with a dichroic and micro lens is shown in Figure~\ref{Efficiency_inluding_microlense}, the efficiency decreases for both channels away from the reference wavelength. The reduction on the blue side is mainly due to the transmission losses of the optical elements, while the decrease on the red side is due to axial color. The latter can be mitigated either by positioning the fiber at the focal plane of the micro lens or by employing aspheric lenses \citep{parvathy}. A local peak is observed near 0.7~µm at a zenith angle of $60^\circ$ as the residual wavelengths act as off-axis field points for the micro lens. Around this wavelength, the corrected rays lie farther from the optical axis, creating a pupil coma \citep{pupil_coma}. This redistribution of rays produces a smaller geometric spot, which increases the throughput.

\subsection{Tolerance analysis}

This section discusses the sources of errors that influence the system throughput, primarily governed by the spot radius. Since HROS is seeing-limited and designed for a stability goal of 1~m/s, the manufacturing requirements are comparatively less stringent than those of adaptive optics (AO)-assisted instruments \citep{Cunningham_2008,simoce,Little_iLocater}.  

For the design up to the camera focal plane in the ADCS, tolerance analysis was performed using the Zemax tolerancing tool, which evaluates surface-related (manufacturing) and element-related (alignment) errors individually. The tool employs a Monte Carlo approach to estimate the impact of perturbations in the optical system. The primary objective of tolerance analysis was to ensure that spot radius variations induced by tolerances remain confined within the 1$''$ fiber footprint specified in the requirements. 

For manufacturing tolerances, we considered the tolerances listed in Table~\ref{tolerance} for each surface in the design, from the TMT focal plane to the camera focus. These included thickness deviations of 0.1~mm, radius deviations of $\pm 1\%$, and x- and y-tilt angle errors of up to $0.5^{\circ}$ \citep{Manju_ADC}. Tilts of the prism surfaces, in particular, affect the effective apex angle. We found that the average image quality decreases by approximately 5\% under the applied surface tolerances.

For alignment tolerances, we applied rotational perturbations of $\pm 0.5^{\circ}$ to the Amici prisms and x- and y-tilts of $1'$, decenter errors of $\pm 1$~mm, as summarised in Table~\ref{tolerance}. These perturbations were introduced one element at a time—collimator, Amici~1, Amici~2, and camera—while keeping the others fixed. On average, the spot diagrams changed by approximately 10\% under these alignment perturbations.  

Although such large decenter values are not typically required during instrument alignment, since they can displace the beam outside the stable fiber and cause significant light loss, they were intentionally adopted in HROS-ADC tolerance analysis. This approach probes the robustness of the design and evaluates image quality under exaggerated alignment errors. By comparison, AO-assisted instruments generally require much tighter tolerance budgets.  

A potential source of concern in high-precision radial velocity spectrographs is the accuracy of melt data, i.e., refractive index inhomogeneities and temperature-dependent variations introduced during glass production. Melt data–related residuals of up to $\sim150$~mas in atmospheric dispersion correction have been reported for the material combination PBL6Y + N-PK51\footnote{Based on reported melt and temperature coefficients, the Nikon~7054–CaF$_2$ pair exhibits refractive-index stability similar to that of the N-PK51–PBL6Y combination. Both pairs show low to moderate sensitivity ($\Delta n \sim 10^{-5}$--$10^{-6}$; $dn/dT \sim 7$--$10\times10^{-6}$~K$^{-1}$). \citep{leviton2015,nikon_i_line_e, schott_data, ohara_pbl6y}.} (0.42–0.78~$\mu$m) when the originally optimized prism apex angles are used. These residuals can be reduced by re-optimizing the prism apex angles \citep{melt}, which is crucial for ultra-stable instruments aiming for $\sim10$~cm/s radial velocity precision. In the HROS-MOS case, with a stability requirement of 1~m/s (corresponding to a tolerance of $\sim1000$~mas), such deviations are not expected to be a dominant limitation. Therefore, melt data were not considered in this analysis.

    \begin{table}
        \centering
        {
        
        \begin{tabular}{ll|ll}
            \hline
            \multicolumn{2}{c}{Manufacturing (Surface)} & \multicolumn{2}{c}{Alignment (Element)} \\
            \hline
            
            X\&Y-Decenter & $\pm$1\,mm & X\&Y-Decenter & $\pm$1\,mm \\
            X\&Y-Tilt & $\pm$1' & X\&Y-Tilt & $\pm$ 1' \\
            Radius & $\pm$1\% & Rotation Angle & $\pm$ 0.5$\degree$ \\
            Thickness & $\pm$ 0.1\,mm & & \\
            
            \hline
        \end{tabular}
        }
        \caption{The table presents the manufacturing and alignment tolerances
considered for the ADC and fore optics design. Manufacturing tolerances include tilts, decenter, radius, and thickness deviations, while the alignment tolerances account for decenter, tilt, and rotational errors of the optical elements. The 2~\arcmin{} of parallelism error is included by assigning a ±1 \arcmin{} alignment tilt tolerance along each axis.}

       \label{tolerance}
    \end{table}

\section{Conclusion}

In this paper, we presented the optical design of an Atmospheric Dispersion Corrector tailored for a broad wavelength coverage, ranging from 0.31 to 1$~\mu\text{m}$. The design mitigates dispersion, bringing throughput at a Zenith angle of 60$^\circ$ across the working wavelength to nearly equal the Zenith angle of 0$^\circ$. This improvement enhances the overall efficiency of the spectrograph while ensuring beam deviations are constrained to keep the light within a 1\arcsec{} fiber. Although the ADCS was specifically designed to meet the requirements of the MOS mode for the HROS on the TMT, the material combination of Nikon 7054 and CaF$_2$, which provides over 95\% transmission for a thickness of 10 mm, renders it suitable for other applications. 

One key part of this work is modifying the atmospheric model within Zemax, enabling a more precise ADC design that effectively addresses dispersion and beam deviations. Following correction, the beam deviations are confined to a minimal area with a radius of only 0.16\arcsec{}, ensuring the required performance. 

To validate the robustness of the design, a tolerance analysis was performed using Zemax to simulate potential manufacturing and alignment errors. The results indicate that while the average image quality decreased by 10\% due to manufacturing and alignment tolerances, this degradation does not significantly impact the throughput to the spectrograph. By convolving the increased spot size due to system perturbations with atmospheric seeing, we demonstrated that the resulting image quality remains within the required specifications.

Furthermore, micro lenses made from Nikon 7054 glass were added to change the system’s focal ratio from an f/15 to $\approx$ f/3 to increase the fiber coupling efficiency.

In conclusion, the proposed ADC design, with its new combination of materials, corrections, and robust tolerance analysis, addresses the stringent requirements of HROS-TMT. Its flexibility and high throughput make it a promising solution for both on-axis and off-axis observations using seeing-limited fibre-fed spectrographs.

\section*{ACKNOWLEDGMENTS} 

We thank the optical engineer, Mr. Bernard Delabre, suggestion to use Nikon glass materials. We also thank the anonymous reviewer(s) for their constructive comments and suggestions, which have helped us improve the quality of this manuscript. We thank the Department of Science and Technology, India TMT team, the Indian Institute of Astrophysics, and the University of Calcutta for their unwavering support throughout this work.

\section*{DATA AVAILABILITY}

The source code and DLL file for the Filippenko 1982 model are included in the GitHub\footnotemark[5] repository. Data information, sourced from the Thorlabs website\footnotemark[8], is also provided. Zemax designs and Zemax-Python interfacing codes will be shared upon request to the corresponding author, with permission from India-TMT.



\bibliographystyle{mnras}
\bibliography{mult_fps}



\bsp	
\label{lastpage}
\end{document}